\documentclass[twocolumn,showpacs,preprintnumbers,amsmath,amssymb,superscriptaddress,bibnotes, prl]{revtex4}

\usepackage{graphicx}
\usepackage{dcolumn}
\usepackage{tabularx}
\usepackage{bm}
\usepackage{braket}
\usepackage{color}
\makeatletter
\def\Ddots{\mathinner{\mkern1mu\raise\p@
\vbox{\kern7\p@\hbox{.}}\mkern2mu
\raise4\p@\hbox{.}\mkern2mu\raise7\p@\hbox{.}\mkern1mu}}
\makeatother

\begin{document}

\preprint{Preprint}

\title{Strong-Field Perspective on High-Harmonic Radiation from Bulk Solids}

\author{Takuya Higuchi} 
\email[E-mail: ]{takuya.higuchi@fau.de}
\affiliation{Department of Physics, Friedrich-Alexander-Universit\"at Erlangen-N\"urnberg, Staudstrasse 1, D-91058 Erlangen, Germany}
\author{Mark I. Stockman} 
\affiliation{Max-Planck-Institut f\"ur Quantenoptik, Hans-Kopfermann-Strasse 1, D-85748 Garching, Germany}
\affiliation{Fakult\"at f\"ur Physik, Ludwig-Maximilians-Universit\"at, Am Coulombwall 1, D-85748 Garching, Germany}
\affiliation{Department of Physics and Astronomy, Georgia State University, Atlanta, Georgia 30303, USA}
\author{Peter Hommelhoff}
\affiliation{Department of Physics, Friedrich-Alexander-Universit\"at Erlangen-N\"urnberg, Staudstrasse 1, D-91058 Erlangen, Germany}
\affiliation{Max-Planck-Institut f\"ur Quantenoptik, Hans-Kopfermann-Strasse 1, D-85748 Garching, Germany}

\date{\today}

\begin{abstract}
Mechanisms of high-harmonic generation from crystals are described by treating the electric field of a laser as a quasi-static strong field.
Under the quasi-static electric field, electrons in periodic potentials form dressed states, known as Wannier-Stark states. The energy differences between the dressed states determine the frequencies of the radiation. 
The radiation yield is determined by the magnitudes of the inter-band and intra-band current matrix elements between the dressed states. The generation of attosecond pulses from solids is predicted.
Ramifications for strong-field physics are discussed.
\end{abstract}
\pacs{72.20.Ht, 42.65.Ky, 42.65.Re}


\maketitle

\begin{figure*}[]
\begin{center}
\includegraphics[width=17.5cm]{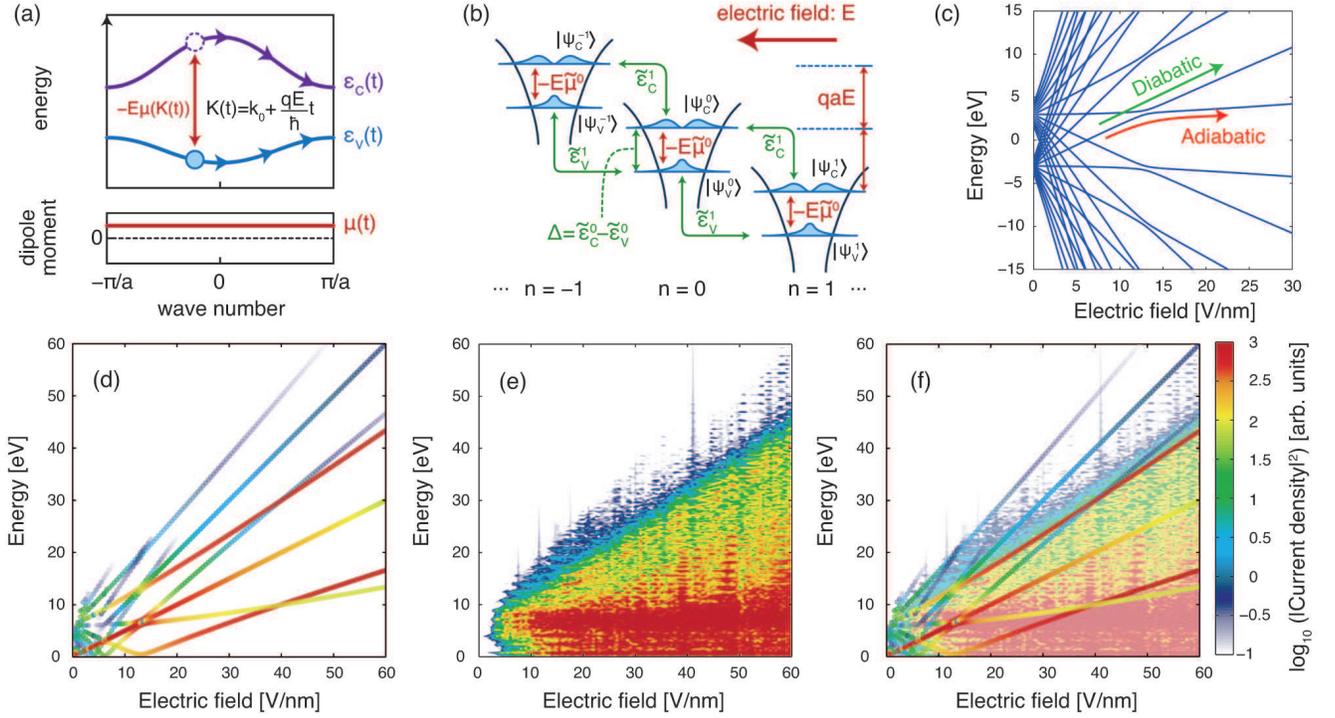}
\caption{\label{Figure1} 
(Color) 
(a) Schematic of the dynamics of electrons in the valence and conduction bands.
The electron experiences both inter-band transition and the intra-band acceleration. 
(b) Coordinate-space representation of the system under a static electric field.
(c) Quasi-energy spectra as functions of the quasi-static electric field. 
(d) Energy spectra of the total current matrix elements as functions of the quasi-static electric field. The color shows the intensity amplitude of the matrix elements. 
(e) Numerically obtained HHG spectra as functions of the peak electric field of the incident laser pulse. 
(d) and (e) are overlapped in (f) to show the correspondence of their features. For better visibility, the color of (e) is made more transparent.}
\end{center}
\end{figure*}

Advances in intense pulsed lasers have opened up an avenue to field-induced non-perturbative nonlinear optical phenomena, such as high-harmonic generation (HHG) and attosecond pulse generation \cite{Corkum:2007ph, RevModPhys.81.163}.
The scope of these strong-field phenomena has been mainly focused on gaseous media, and was extended to solids more recently \cite{Ghimire:2011fk, SchubertO.:2014uq, Schiffrin:2013rc, Ivanov20133, Krausz:2014dn, PhysRevLett.105.086803, PhysRevLett.107.086602, PhysRevB.87.115201}. In particular, HHG from wide band gap semiconductors under illumination of low-frequency laser light has been reported \cite{Ghimire:2011fk, SchubertO.:2014uq}, and the door to extreme wavelength conversion employing condensed-matter materials was opened.

The non-perturbative character of the HHG manifests itself as plateau structures in its energy spectrum.
The cutoff energy of the plateau provides insights into the electronic dynamics on attosecond timescales.
In the case of HHG from atoms, the HHG process is well described by the three step model: field ionization, acceleration, and recollision \cite{PhysRevLett.71.1994, PhysRevA.49.2117}. The cutoff energy of the resultant radiation is given by the maximal kinetic energy gained by the electron during excursion. It scales quadratically with the field amplitude, on top of the ionization potential of the atom. 
A similar insightful model for HHG from solids is desired but has so far been elusive, and the three-step model is not applicable to solids, as is evidenced by the linear scaling of the experimentally observed cutoff energies with the peak field strength \cite{Ghimire:2011fk, SchubertO.:2014uq}.

The dynamics of electrons in solids has been analysed by numerical simulations based on the integration of the time-dependent Schr\"odinger equation (TDSE). In particular, the interplay of inter-band and intra-band interactions has been shown to play an important role to determine the HHG radiation spectrum \cite{PhysRevB.77.075330, 1367-2630-15-2-023003, 1367-2630-15-1-013006}.
Excellent agreement of the numerical simulations with the experimental results confirms the validity of this approach \cite{SchubertO.:2014uq}.
Various proposals based on such numerical simulations have been made, for example isolating an attosecond pulse by using two-color laser pulses \cite{PhysRevB.84.081202}.

These numerical simulations, however, require further interpretation of the results, and a more direct way of determining the HHG cutoff energy has been eagerly demanded \cite{Ivanov20133}.
Various methods to determine the cutoff energies have been proposed,
but they cannot fully explain the observed cutoff energies of HHG from solids, as discussed in what follows.

(1) Ghimire {\it et al.} considered the intraband current as the source of HHG \cite{Ghimire:2011fk, PhysRevA.85.043836}. Due to the non-parabolicity of the conduction-band energy, the intraband current contains harmonics of the Bloch frequency $\Omega_{\rm B} \equiv qaE/\hbar$, where $q$ is the unit charge, $a$ is lattice constant of the crystal, and $E$ is the electric field amplitude of the laser.
This model explains the linear scaling of the cutoff energy with the laser field amplitude, but cannot treat the additional offset observed in the cutoff energy. 

(2) Another approach is to consider the frequency of the oscillating inter-band polarization \cite{Ivanov20133, Vampa:13}. In this model, highest-energy photons are emitted at the top of the bands: this gives the upper limit of the HHG energy, but does not explain the linear scaling, and the experimentally obtained cutoff \cite{Ghimire:2011fk, SchubertO.:2014uq} was larger than this limit.

In these models, inter- and intra-band contributions of the light-matter interactions are considered separately.
This separation is allowed only when the field is weak enough to be treated as a perturbation, where the electronic band states stay as a good basis set. 
When the electric field is strong and the light-matter interaction is non-perturbative, however, the interplay between the inter- and intra-band contributions cannot be neglected \cite{PhysRevB.77.075330, 1367-2630-15-2-023003, 1367-2630-15-1-013006}, and they should be considered together.
This artificial separation hinders proper prediction of the HHG radiation frequency.
Otherwise, both the population and the matrix elements contains oscillatory components, in analogue to the carrier-wave Rabi flopping in two-level systems \cite{Plaja:92, PhysRevA.62.055401, Golde:06}. 

In this study, we propose a simple analytical model to gain insight into the physical processes involved in the HHG radiation in solids, and to obtain the cutoff energies.
The essential point of our model is to consider electronic states dressed with a quasi static electric field, which are known as Wannier-Stark localised states \cite{PhysRevB.36.7353, PhysRevLett.105.086803, PhysRevLett.107.086602}. Both inter- and intra-band couplings are simultaneously treated.
The energy differences between these dressed states give the photon energy of the radiation, 
while the radiation yield is determined by the magnitude of the current matrix element between them.
We find excellent agreement between this new method and numerical simulations based on solving the TDSE.
The experimentally observed cutoff energy is well understood on the basis of this model, including both the linear scaling with the field amplitude as well as the offset.
In addition, we show that the highest energy photons from solids are emitted when the laser field 
is the strongest, which suggests participation of the adiabatic WS states.
Based on this understanding, we predict the possibility of generation of isolated attosecond pulses from solids.

The optical electric field interacts with electrons in a periodic lattice
through two processes, the inter-band and intra-band couplings, which are found in the Schr\"odinger equation for semiconductors \cite{Golde:06, PhysRevB.77.075330, SchubertO.:2014uq}:
\begin{eqnarray}
&&  {\cal H}(t) = 
\int_0^{\frac{2\pi}{a}} dk \bigg[
\sum_{\lambda} \varepsilon _{\lambda}(k) \hat a^\dagger _{\lambda,k}\hat a _{\lambda,k} \nonumber \\
&-& E(t) \Big [\sum_{\lambda, \lambda'} \mu_{\lambda\lambda'}(k) \hat a^\dagger _{\lambda,k} \hat a _{\lambda',k} 
+ i q \sum_{\lambda} \hat a^\dagger _{\lambda,k} \nabla_{k} \hat a _{\lambda,k} 
\Big]
\bigg]. \label{eq-SOBE}
\end{eqnarray}
Here $\hat a_{\lambda,k}$ is the annihilation operator of an electron with a wave number $k$, 
the indices $\lambda$ and $\lambda'$ label the bands.
$\varepsilon_\lambda(k)$ is the electron energy of band $\lambda$ at wavenumber $k$,
$\mu_{\lambda\lambda'}(k)$ is the inter-band dipole moment between the bands $\lambda$ and $\lambda'$ for wavenumber $k$.
The three terms describe the energies of the band electrons, the inter-band polarization, and the intra-band polarization, respectively.

We are interested in higher-harmonic radiation, which oscillates much faster than the electric field. Therefore, we assume that the laser electric field is quasi-static, $E(t)=E_0$, and watch the electronic states under this assumption. 
The last term in Eq.~\eqref{eq-SOBE} results in the acceleration theorem describing the intra-band motion of the electrons: $K(t) = k_0 +\frac{qE}{\hbar}t$, where $k_0$ is the initial wave number, as schematically depicted in Fig.~\ref{Figure1}(a).
The electron-electron interaction works as pure dephasing at the time scale of tens of femtoseconds \cite{Golde:06}, so the dynamics of the electrons with different initial $k$ values can be independently calculated.
We take an interaction-representation picture for $k$ so that $k=K(t)$ changes in time following this intra-band acceleration. So the last term in Eq.~\eqref{eq-SOBE} is eliminated in this picture.
The cost we pay is that the Hamiltonian now depends on time even though we assumed a static electric field.
However, this temporal dependence is periodic in time, hence we can employ the Floquet theorem to obtain the solutions.

We illustrate how to apply the Floquet method using a two-band model for example, i.e., conduction band and valence band ($\lambda = $ C or V, respectively) $\mu_{{\rm C},{\rm V}}(k) = \mu^*_{{\rm V},{\rm C}}(k) \equiv \mu(k)$. Note that this procedure is actually valid for arbitrary number of bands.
In matrix form, the Hamiltonian is
\begin{equation}
H(t) = 
\left[
\begin{array}{cc}
\varepsilon _{\rm V}(K(t)) & -E_0 \mu (K(t)) \\
-E_0 \mu^* (K(t)) & \varepsilon _{\rm C}(K(t))  \\
\end{array}
\right]. \label{eq-OriginalHamiltonian}
\end{equation}
The conduction- and valence-band energies and the dipole coupling energies are periodic functions of $k$.
Under a static field, this periodicity is imprinted onto temporal periodicity because $K(t)$ is linear in $t$.
Therefore, the Hamiltonian is also periodic in time, and can be decomposed into a Fourier series as
\begin{equation}
H(t) = \sum _n e^{-i n \Omega_{\rm B} t} \tilde H^n,~~ 
\tilde H^n 
\equiv  \left[  
\begin{array}{cc}
\tilde \varepsilon _{\rm V}^n & -E_0 \tilde \mu^n  \\
-E_0 \tilde \mu^{n*} & \tilde \varepsilon _{\rm C}^n \\
\end{array}
\right].
\end{equation}
where $\Omega_{\rm B} = aeE_0 / \hbar$ is the frequency of the periodicity,
which is the Bloch frequency.
$\tilde \mu^{n}$ and $\tilde \varepsilon_{\lambda}^n$ are the Fourier coefficients of the dipole moment and the band energies, respectively.

We can now apply the Floquet theorem to this system.
The problem to find solutions of the original Schr\"odinger equation \eqref{eq-OriginalHamiltonian} is translated into solving the following eigenvalue problem \cite{Chu20041}:
\begin{equation}
\sum_{\nu',n'} \left( H_{\nu\nu'}^{n-n'} - n \hbar \Omega_{\rm B} \delta_{\nu \nu'}\delta^{nn'} \right) \ket{\phi_{\nu'}^{n'}} 
= \epsilon_ {\nu}^n \ket{\phi_{\nu}^n}. 
\end{equation}
$\epsilon_ {\nu}^n$ is a quasi-energy, and the eigenstate $\ket{\phi_{\nu}^n}$ is a Wannier-Stark state \cite{Gluck:1998ak}.
Here the indices $\nu$ and $\nu'$ are the labels of different Floquet quasi-energy series. Within one series, the quasi-energies are harmonic (equidistant spectrum), i.e., $\epsilon_{\nu}^{n}-\epsilon_{\nu}^{n'}=\hbar (n-n') \Omega_{\rm B}$.
The number of the series is the same as the number of the original electronic bands.
The solutions of the original Schr\"odinger equation are constructed from the quasi-energy eigenstates of the Floquet Hamiltonian:
\begin{equation}
\ket{\Psi (t)} = \sum_\nu  \sum _{n} C_{\nu} e^ {-i \frac{\epsilon_\nu^n}{\hbar} t } \ket{\phi_{\nu}^{n}}.
\label{eq-FloquetState}
\end{equation}
The linear-combination coefficients $C_{\nu}$ are determined by the initial electron states and the prior temporal evolution of the field, following the Landau-Zener tunnelling probability \cite{LandauTunneling, Zener01091932,PhysRevLett.105.086803, Schiffrin:2013rc}

It is interesting to see that the above procedure of the Floquet method can be mapped into a coordinate-space picture, as shown in Fig.~\ref{Figure1}(b).
For this, we start from a standard tight-binding procedure. Namely, consider atomic states $\ket{\phi_{\lambda}^n}$ in a lattice, where $n$ indicates the position of the atomic site and $\lambda$ is the band index.
These atomic states couple to each other. 
Without electric field,
$\tilde \varepsilon _{\lambda}^n$ corresponds to the coupling to the $n$-th neighbor site.
Diagonalization of the Hamiltonian for this condition gives the standard band structure.
An external electric field provides two additional effects.
The first is the position-dependent energy shift because the electric field produces a potential, $-qxE$, where $x$ is the position. 
The second is the dipole coupling $\tilde \mu^n$ that causes inter-band mixing between the $n$-th neighbor sites.
Diagonalization of this coordinate-space Hamiltonian gives the same energy spectrum as the quasi-energy spectrum in the Floquet analysis.

Figure \ref{Figure1}(c) shows the quasi-energy spectra as function of the quasi-static field amplitude.
Parameters are chosen to simulate a typical wide band-gap semiconductor:
a band offset $\Delta \equiv \tilde \varepsilon_{\rm C}^0 - \tilde \varepsilon_{\rm V}^0$ of $6$~eV,
a conduction-band width $2 \tilde \varepsilon_{\rm C}^1=2 \tilde \varepsilon_{\rm C}^{-1}$ of $3$~eV,
a valence-band width $2 \tilde \varepsilon_{\rm V}^1=2 \tilde \varepsilon_{\rm V}^{-1}$ of $2$~eV, 
and an intra-atomic dipole moment $\tilde \mu^0$ of $0.1$~$|e^-|\cdot$nm.
The other parameters ($\tilde \varepsilon_{\lambda}^n$ for $|n|\geq 2$ and $\tilde \mu^n$ for $|n|\geq 1$) are zero.
For the diagonalization, we introduced a cutoff in $n$ as $|n|\leq 7$. Increasing the cutoff does not change the quasi-energies of the $n=0$ states for $\Omega_{\rm B}>\Delta$ (i.e., $|E|>\Delta/qa$) because the mixing between wave functions having a large difference in $n$ is negligibly small.
In the coordinate space picture [Fig.~\ref{Figure1}(b)], this corresponds to neglecting the inter-atomic coupling if they are separated by $na$, which is larger than the Wannier-Stark localization length \cite{PhysRevLett.105.086803}.
Note it is also possible to diagonalize the Floquet Hamiltonian fully analytically by means of infinite-variable Bessel functions \cite{0305-4470-35-9-101}.

Next, we calculate the current.
The intra-band current operator is obtained from the electron group velocity as 
\begin{equation}
J_{\lambda\lambda'}(t) = \frac{e}{\hbar} \frac{\partial \varepsilon_{\lambda}(k)}{\partial k}\Big|_{k=K(t)} \delta_{\lambda\lambda'}.
\end{equation}
The inter-band current is given as the temporal derivative of the interband polarization
\begin{equation}
P_{\lambda\lambda'}(t) = \mu_{\lambda\lambda'}(K(t)).
\end{equation}
Both are periodic in time, and thus can be described using their Floquet Matrix elements.
We calculate the expectation value of the total current for the dressed electronic state in Eq.~\eqref{eq-FloquetState}:
\begin{eqnarray}
 &&\frac{d}{dt} \braket{P(t)} +
\braket{J(t)} \nonumber = \sum _{\nu,\nu'} 
\sum _{n,n'} 
C_{\nu}^*C_{\nu'} 
e^ {i\frac{\varepsilon_{\nu}^{n} - \varepsilon_{\nu'}^{n'}}{\hbar} t } \\
&& \times \left(
\bra{\phi_{\nu}^{n}} 
P_{\rm F}
\ket{\phi_{\nu'}^{n'}} 
\frac{i(\varepsilon_{\nu}^{n} - \varepsilon_{\nu'}^{n'})}{\hbar}
+
\bra{\phi_{\nu}^{n}} 
J_{\rm F}
\ket{\phi_{\nu'}^{n'}} 
\right). \label{eq-CurrentExpectation}
\end{eqnarray}
Here, the Floquet matrix $P_{\rm F}$ is defined as 
$\bra{\psi^n_{\lambda}} P_{\rm F} \ket{\psi^{n'}_{\lambda'}} 
\equiv \bra{\psi_{\lambda}} \tilde P^{(n-n')} \ket{\psi_{\lambda'}}$,
where 
$\ket{\psi^{n}_{\lambda}}$ are the bases in the extended Hilbert space and
$\ket{\psi_{\lambda}}$ are the bases of the original equation \eqref{eq-OriginalHamiltonian}.
$J_{\rm F}$ is defined similarly. See supplementary information for details.

Equation~\eqref{eq-CurrentExpectation} predicts the cutoff energies and the radiation yield.
The difference between quasi-energy states, $\varepsilon_{\nu}^{n} - \varepsilon_{\nu'}^{n'}$, gives the photon energy of the radiation.
Figure~\ref{Figure1}(c) shows energy spectra of the radiation. 
According to the quasi-energy spectrum, seemingly infinitely high energy photons can be emitted because the quasi-energy spreads over infinite values. 
However, this is not the case because the term in parenthesis in Eq.~\eqref{eq-CurrentExpectation} between different quasi-energy eigenstates steeply drops as the quasi-energy difference increases. 
This is encoded in the color in Fig.~\ref{Figure1}(d).
Note the coefficients $C_{\nu}$ reflect the population distributions among the different bands,
but this affects the population within a factor, while the value of the matrix elements show variations of orders of magnitudes, which we will see later.

To show the validity of this picture, we compare it with numerical results, as shown in Fig.~\ref{Figure1}(e).
The temporal evolution of the Schr\"odingier equation is obtained with the Crank-Nikolson method \cite{PSP:2035556}.
The valence band is initially fully occupied, while the conduction band is empty.
We calculate the total current $\frac{d}{dt} \braket{P(t)} + \braket{J(t)}$.
The laser pulse has a central frequency of 200 THz and a FWHM of the intensity envelope of 30 fs.
We changed the peak electric field in the simulations, while fixing the waveform.
Carrier-envelope-phase variation induces negligible change in HHG spectra for such relatively long pulses.
The numerical results show several cutoff steps in the energy (change in color), which clearly coincide with the radiation yield spectra obtained under the quasi-static field assumption, as shown in Fig.~\ref{Figure1}(f).
Hence, the analytical results are well supported by numerical simulations.

\begin{figure}[t]
\begin{center}
\includegraphics[width=8.5cm]{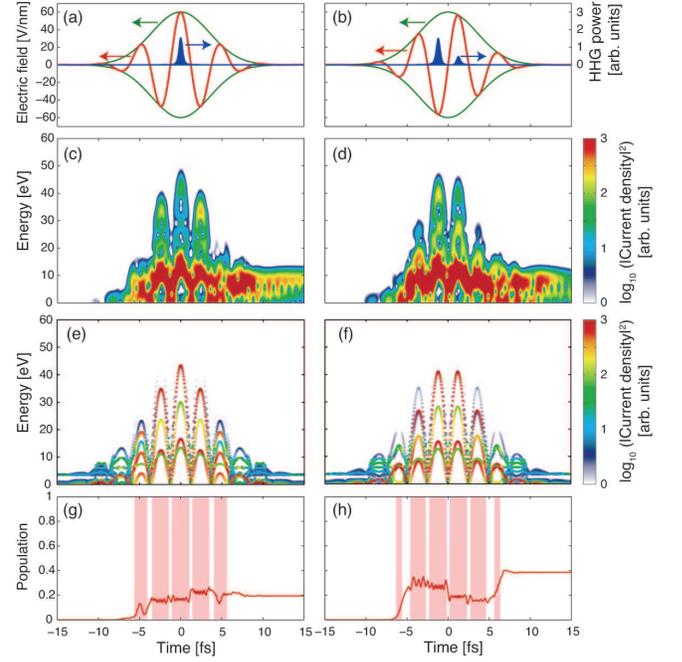}
\caption{\label{Figure2} 
(Color) Prediction of the possible generation of attosecond pulse. The incident laser waveforms having (a) cosine and (b) sine wave forms are plotted as the red curves, while their envelopes (green curves) are identical. The blue shaded areas show the power of the HHG through high-pass filters, having cutoff energies at (a) 40 eV and (b) 38 eV. 
(c)(d) Numerically obtained HHG spectrograms. The window function is a gaussian having a FWHM of 0.67 fs, i.e., 6.2 eV in energy, which broadens the spectrograms.
(e)(f) Spectra of the total current matrix elements as functions of time. The color indicates the intensity amplitude of the matrix elements.
(g)(h) Temporal evolution of the upper-level population. The shaded areas indicate $|\hbar \Omega_{\rm B}|>\Delta$, indicating the field amplitude exceeds the last anti-crossing in Fig.~\ref{Figure1}(b).}
\end{center}
\end{figure}

So far we have clarified the relationship between the quasi-static electric field and the cutoff energy.
This provides us an opportunity to predict the possibility of generation of isolated attosecond pulse(s), whose waveforms can be controlled by the carrier-envelope phase (CEP) of the incident laser pulse.
Figures \ref{Figure2}(a) and (b) show two initial laser waveforms having different (0 and $\pi/2$) CEPs, which correspond to cosine and sine waveforms, respectively.
We numerically simulate the temporal evolution of the TDSE under these laser waveforms, and obtain spectrograms of the currents [Figs.~\ref{Figure2}(c) and (d)]. These spectrograms show notable differences. The cosine pulse generates a single high-energy peak while the sine pulse generates a double peak. 
Single or double attosecond pulses can be separated from the rest of the radiation by introducing high-pass filters, as shown in 
Figs.~\ref{Figure2}(a) and (b).

The quasi-static assumption insightfully accounts for the main features in the spectrograms when $|\hbar\Omega_{\rm B}|>\Delta$. Figures~\ref{Figure2} (e) and (f) show the energy and the yield of the radiation under the quasi-static field approximation,
which well explains the photon energies of the radiation peaks in Figs.~\ref{Figure2}(c) and (d).
Note that the window function for the wavelet analysis has a spectral window of $6.2$ eV, which broadens the spectrograms. Also, after the laser pulse passed, there remains a considerable radiation at low energies in the spectrogram. This cannot be treated in the quasi-static field model because it is applicable when $|E|>\Delta/qa$.

One limitation of the present approach is that the linear combination coefficients $C_{\nu}$ cannot be determined with the present value of the field alone, because they are determined by the initial conditions and depend on how the field evolved. Therefore, it is worth considering how these coefficients evolve in the numerical simulations.
The speed of the change in the field value determines the tunnelling rate when the field value goes through anti-crossings in the quasi-energy spectra. 
For example, the last anti-crossing in Fig.~\ref{Figure1}(c) is $\sim 0.5$ eV wide (i.e., the one for the largest field amplitude), which is comparable to the frequency of the laser field, $0.83$ eV/$\hbar$. Therefore, when the field value goes across the anti-crossings, electrons experience intermediate transitions between adiabatic and diabatic ones through Landau-Zener tunnelling \cite{PhysRevLett.105.086803, Schiffrin:2013rc, LandauTunneling, Zener01091932}. This is found in Figs.~\ref{Figure2} (g) and (h), which show the populations of the upper energy level on an atomic site.
This temporal evolution of the populations accounts for the more detailed features in the HHG spectrograms, which are not explained by the magnitudes of the matrix elements alone.
For example, the two high-energy pulses in Fig.~\ref{Figure2}(d) have different intensities, and this difference reflects the difference in upper-level population in Fig.~\ref{Figure2}(h). The evolution of the upper-level population is important to understand other strong-field phenomena, such as laser-field induced currents in dielectrics \cite{Schiffrin:2013rc}. The radiation yield obtained with the present method can bridge the gap between these intriguing phenomena.

Our method is applicable to any one-dimensional band structure, and can treat any number of bands.
Extensions of this model to higher dimensions will be published else where.
Band structures and the dipole moments are periodic functions of the wave vector.
When the quasi-static electric field is not along the crystalline axis, the Hamiltonian has multiple periods. In this case, one possibility is to employ the many-mode Floquet theory \cite{Ho83a}. Such multiple periodicities could induce dephasing of the electrons due to diffraction into different directions. Therefore, it is worth extending the current scheme into larger dimension and compare it with experiments and numerical simulations. 

To summerize, energies and wave functions of the field-dressed states are important to determine the HHG cutoff from solids when the field exceeds the critical strength, $|E|>\Delta/qa$.
The differences of the quasi-energies of the Wannier-Stark states determine the radiation energies.
The radiation yields are determined by the current matrix elements between different quasi-energy states.
This radiation mechanism is analogous to the one employed in the quantum cascade laser, where the mini-bands are formed in semiconductor superlattices under a static field, and the radiation frequency corresponds to the energy difference between mini-bands \cite{Faist22041994, Williams:2007fh}. In this sense, HHG in solids can be considered as quantum-cascade emission at extreme ultraviolet frequencies, where high-energy carriers are coherently injected through Landau-Zener tunnelling.
Highest energy radiation is emitted when the incident field peaks. This greatly differs from the atomic case, where the recollision event of the electrons with highest kinetic energy does not happen at the time when the laser field peaks.

The authors acknowledge S. C. Furuya for helpful discussions.
This work was supported by the DFG Cluster of Excellence Munich-Centre for Advanced Photonics
and ERC grant ``Near Field Atto.''
TH acknowledges support by JSPS postdoctoral fellowship for research abroad.
Major funding for MIS was provided by Grant No.~DE-FG02-01ER15213 from the Chemical Sciences, 
Biosciences and Geosciences Division. Supplementary funding came from Grant No.~DE-FG02-11ER46789 from the Materials Sciences and Engineering Division of the Office of the Basic Energy Sciences, Office of Science, U.S. Department of Energy, Grant MURI No.~N00014-13-1-0649 from the US Office of Naval Research.


\end{document}